\documentclass{statsoc}

\usepackage[a4paper]{geometry}
\usepackage{graphicx,subfigure}
\usepackage[textwidth=8em,textsize=small]{todonotes}
\usepackage{amsmath,microtype,animate,url}
\usepackage{natbib}
\usepackage[pdftex,colorlinks=true]{hyperref}
\hypersetup{citecolor=darkblue,linkcolor=darkblue,urlcolor=darkblue}
\definecolor{darkblue}{rgb}{0,0,.6}

\graphicspath{{plots/}}

\title[Visualising rate of change]{Visualising rate of change: an application to age-specific fertility rates}
\author[H. L. Shang]{Han Lin Shang}
\address{Australian National University, Canberra, Australia}
\email{hanlin.shang@anu.edu.au}

\begin{document}
\begin{abstract}
Visualisation methods help in the discovery of characteristics that might not have been apparent using mathematical models and summary statistics. However, visualisation methods have not received much attention in demography, with the exceptions of scatter plot and Lexis surface. We utilise a phase-plane plot to visualise the rate of change, obtained from derivatives of a continuous function. The phase-plane plot bears a resemblance to hysteresis loops, isogrowth curves, and solutions to differential equations. Using Australian and Chilean fertility, we present phase-plane plots. Similarly to the scatter plot and Lexis surface, the phase-plane plot identifies the age with maximum fertility rate and displays skewness of fertility distribution. Unlike the scatter plot and Lexis surface, the phase-plane plot identifies the age with maximum positive or negative velocity (i.e., trend), can compare the magnitude of the rate of change between any two years based on the size of the radius of circles. The phase-plane plot allows the visualisation of dynamic changes in fertility for a given age over the years and is potentially useful for visualising dynamic changes in birth-cohort fertility. Via the animate package in \LaTeX, a dynamic phase-plane plot is also proposed to visualise changes in fertility over age or year. 
\\

\end{abstract}

\section{Introduction}

Methods for analysing age-specific demographic data can be loosely divided into two categories: those considering age a discrete variable in real space $R^p$ ($p$ represents the total number of ages) or those considering age a continuous variable in function space bounded within a finite interval. For methods considering age a discrete variable, \cite{Bell92}, \cite{Lee93}, \cite{LC92}, and \cite{LL05} have performed extensive work on modelling and forecasting age-specific fertility and mortality, where the demographic rates at different ages are considered discrete data points. For methods considering age a continuous variable, \cite{DPR11}, \cite{HBY13}, \cite{HU07}, and \cite{Shang16} have introduced functional data analysis to model and forecast age-specific demographic rates at a given year as a continuous and smooth function, which can later be converted to discrete ages at any sampling interval. \cite{HU07} emphasised that the main advantage of functional data analysis is that the smoothing technique can be incorporated into the modelling and forecasting of functions, so any missing or noisy data can be naturally dealt with.

Aside from incorporating smoothness into modelling and forecasting, the ability to consider derivatives, a by-product of conceiving of the data as functions, is also of great advantage for visualisation. The first derivative, also known as `velocity' in physics, measures the rate at which the value of a function $f_t(x)$ changes about the change of a continuous variable $x$ or $t$. Here, $x$ denotes age variable, and $t$ denotes time variable, both of which are continuous. The second derivative, also known as `acceleration' in physics, measures the rate at which the rate of change alters the change of variable. For example, \citet[][pp. 86-89]{RS02} considered the human-growth curve. When a child grows more quickly, velocity increases rapidly, decreasing when growth slows. When a child begins to grow, the absolute acceleration (i.e., growing potential) is at its maximum, before gradually decreasing during the rapid growing period. Then, absolute acceleration begins to increase during the slow-growing period, before stabilising when the child stops growing. Velocity and acceleration, which divide two phases into four, are neither visible in scatter plot nor in Lexis surface. Velocity and acceleration both possess important information on rate of change. Thus, analysing velocity and acceleration information may reveal patterns that are not easily seen in the original functions. We aim to extend the phase-plane plot from human-growth curves to demographic functions. Using the Australian and Chilean age-specific fertility rates described in Section~\ref{sec:2}, we consider phase-plane plots for visualising the rate of change of age-specific fertility rates in Section~\ref{sec:3}. In Section~\ref{sec:4}, we consider phase-plane plots for visualising the rate of change of birth-cohort fertility rates. Section~\ref{sec:5} concludes. The computational code in \textsf{R} is available in the supplementary material.

\section{Data sets}\label{sec:2}

\begin{figure}[!ht]
\centering
\subfigure[Rainbow plot]
{\includegraphics[width=12.2cm]{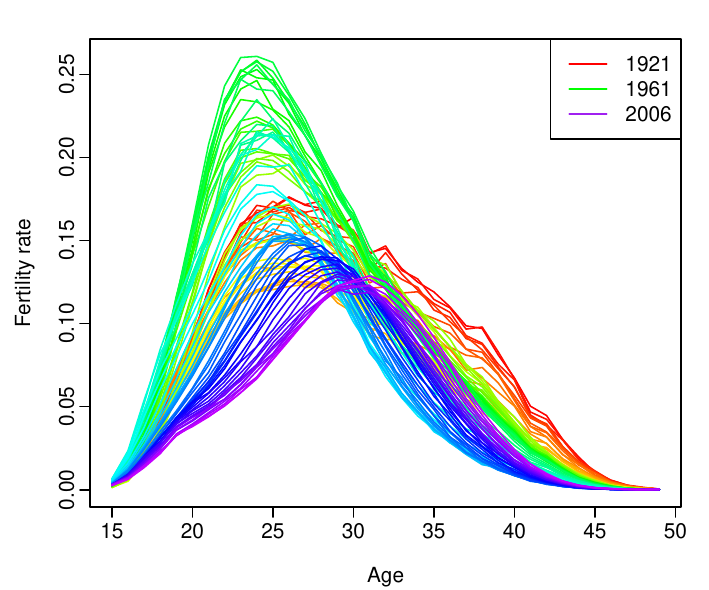}\label{fig:1a}}
\qquad
\subfigure[Lexis surface]
{\includegraphics[width=12.2cm]{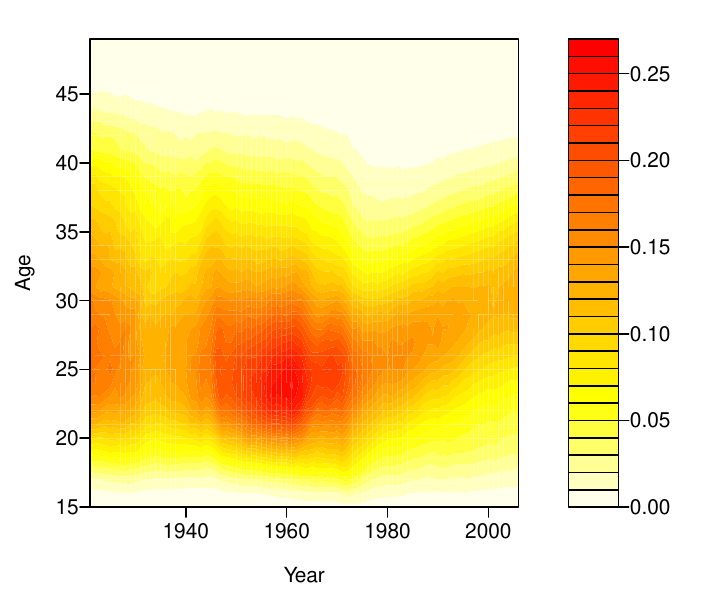}\label{fig:1b}}
\caption{Plots of observed age-specific fertility rates for Australia from 1921 to 2006. Source: Australian Bureau of Statistics (Cat. No. 3105.0.65.001, Table 38).}\label{fig:5_AUS}
\end{figure}

We consider annual age-specific Australian fertility rates from 1921 to 2006. The data set was obtained from the Australian Bureau of Statistics (Cat. No. 3105.0.65.001, Table 38) and is available in the \textit{rainbow} package \citep{SH16} in \textsf{R} \citep{Team16}. The data consist of annual fertility rates by the age of the mother in single years from ages 15 to 49. Graphical displays of the data are given in Fig.~\ref{fig:5_AUS}. The rainbow plot in Fig.~\ref{fig:1a} reveals the phenomenon of recent fertility postponement (shown in purple colour). The Lexis surface in Fig.~\ref{fig:1b} shows the increases in fertility rates between ages 20 and 30 from 1940 to 1980, and this reflects the baby boom period occurred at approximately 1960 \citep[see, e.g.,][]{CCB+88}.

\begin{figure}[!ht]
\centering
\subfigure[Rainbow plot]
{\includegraphics[width=12.2cm]{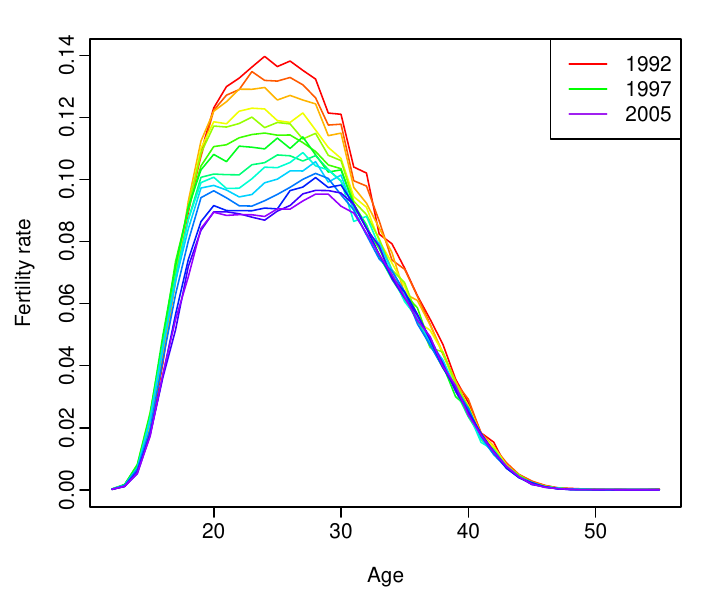}\label{fig:1a_CHL}}
\qquad
\subfigure[Lexis surface]
{\includegraphics[width=12.2cm]{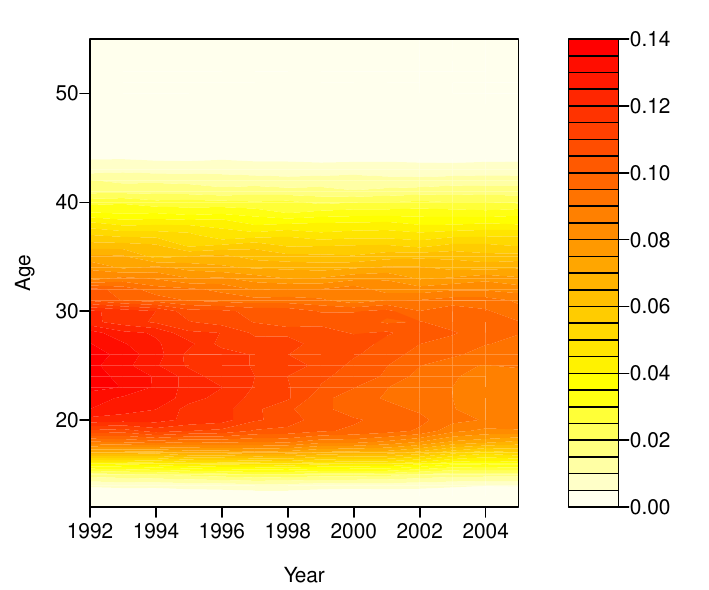}\label{fig:1b_CHL}}
\caption{Plots of observed age-specific fertility rates for Chile from 1992 to 2005. Source: \cite{HFD17}.}\label{fig:1_CHL}
\end{figure}

To analyse contemporary bimodal fertility curves in Latin America, we also consider annual age-specific Chilean fertility rates from 1992 to 2005. The data set was obtained from the \cite{HFD17}. The data consist of yearly fertility rates by the age of the mother in single years from ages 12 to 55. The fertility rate at age 12 includes all fertility rates at or below 12 years of age, while the fertility rate at age 55 consists of all fertility rates at or above 55 years of age. Graphical displays of the data are provided in Fig.~\ref{fig:1_CHL}. The rainbow plot in Fig.~\ref{fig:1a_CHL} reveals the phenomenon of fertility bimodality in recent years (shown in purple colour). The Lexis surface in Fig~\ref{fig:1b_CHL} shows the decreases in fertility rates between ages 16 and 38 from 1992 to 2005, and this may be due to improvements in standards of education and other economic and social conditions \citep{AM11}. 

\section{Visualising rate of change: phase-plane plot}\label{sec:3}

Derivatives are useful in visualising functional data, an example of this is age-specific fertility rates observed over a period. To visualise the rate of change in age-specific fertility rates, we consider the phase-plane plot in functional data analysis \citep[][Section 3.3]{RS02}. The phase-plane plot is a two-dimensional scatter plot of the acceleration against the velocity of a function. In the context of age-specific fertility rate, such a continuous function can be an age or time variable. For example, the velocity and acceleration functions about age $x$ can be expressed as
\begin{align*}
\text{velocity} = \frac{df_t(x)}{dx},\qquad \text{acceleration} = \frac{d}{dx}\left[\frac{df_t(x)}{dx}\right].
\end{align*}

\begin{figure}[!ht]
  \centering
\includegraphics[width=\textwidth]{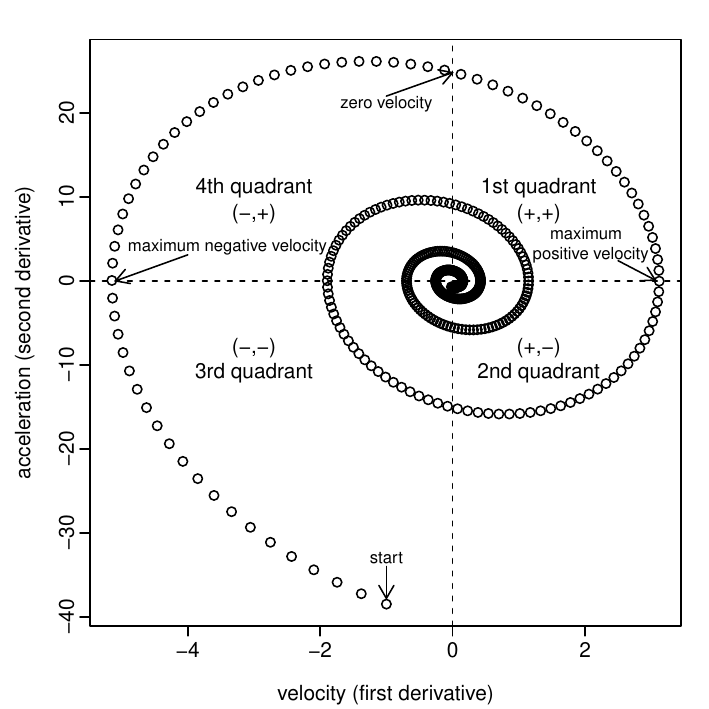}
  \caption{A two-dimensional scatter plot of the acceleration against the velocity for a damped sine function $f(x)=\exp^{-x}\cos(2\pi x)$, where $x=0, 0.01, \dots, 4$.}\label{fig:1}
\end{figure}

For ease of interpretation, we consider a damped sine function $\exp^{-x}\cos(2\pi x)$, which has the first derivative $-\exp^{-x}[2\pi\sin(2\pi x)+\cos(2\pi x)]$ and second derivative $\exp^{-x}[4\pi\sin(2\pi x)+(1-4\pi^2)\cos(2\pi x)]$, for $x=0,0.01,\dots,4$. A phase-plane plot is presented in Fig.~\ref{fig:1}, where the highest absolute velocity or acceleration is reached when the other variable is zero.

From the highest absolute acceleration to the highest absolute velocity (first and third quadrants), the phase-plane plot displays an increase in the rate of change. From the highest absolute velocity to the highest absolute acceleration (second and fourth quadrants), the phase-plane plot shows that the rate of change not only decreases but also reaches zero. When the velocity (i.e., the trend of fertility rates) reaching zero, it indicates the occurrence of maximum or minimum fertility. As noted by \citet[][pp. 30-33]{RS06}, the size of the radius of circles can imply the magnitude of the rate of change; that is, the greater the radius, the greater the magnitude of the rate of change. 

\subsection{Comparison of age-specific fertility rates}

\subsubsection{Australian fertility rates}

\begin{figure}[!ht]
  \centering
  \subfigure[Phase-plane plot]
{\includegraphics[width=10.5cm]{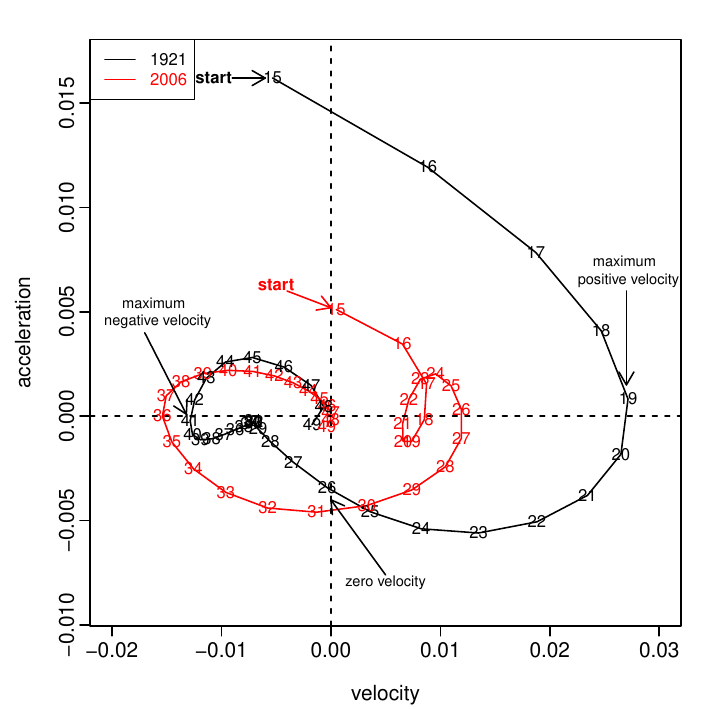}\label{fig:2a}}
\qquad
  \subfigure[Scatter plot]
{\includegraphics[width=10.5cm]{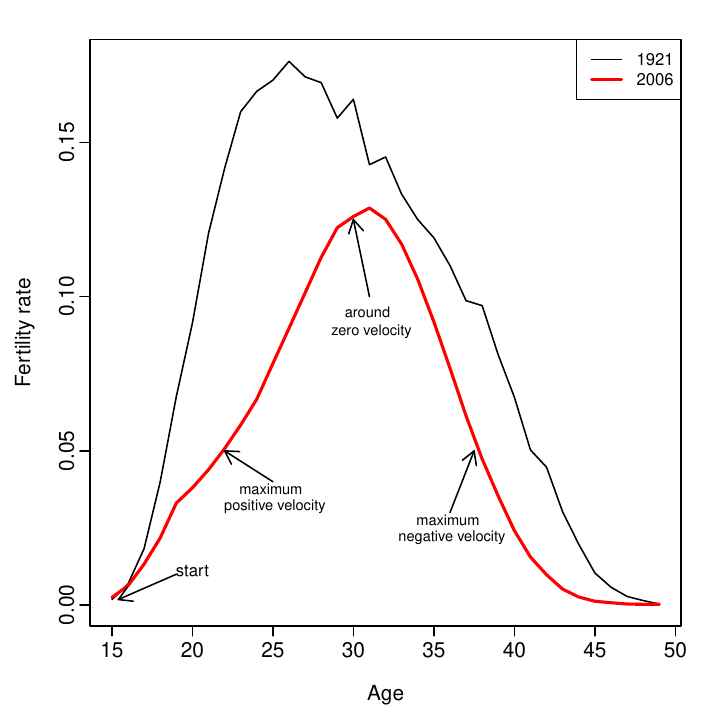}\label{fig:2b}}
  \caption{Phase-plane plot and scatter plot of the age-specific fertility rates for Australia in 1921 (as shown by black colour) and 2006 (as shown by red colour).}\label{fig:2}
\end{figure}

The phase-plane plot allows the visualisation of changes in the shapes of the cycles from year to year. From these changes in shapes, we aim to reveal social changes in age-specific fertility rates, including the phenomena of the baby boom that occurred at approximately 1960 \citep{CCB+88} and the fertility postponement that occurred in more recent years \citep{McDonald00}. In Fig.~\ref{fig:2}, we display the phase-plane plot and scatter plot for age-specific fertility rates at the first year of data in 1921 and the last year of data in 2006 to highlight the specific contrasts between these years in the age distribution of fertility rates.

Fig.~\ref{fig:2a} reveals that the age-specific fertility rates in 1921 increase in velocity from ages 15 to 19, reaching the highest positive velocity at age 19, before showing a decreasing velocity from ages 20 to 26. Maximum fertility occurs at age 26, as shown by zero velocity. The absolute velocity increases slowly from ages 27 to 40 and reaches the maximum negative velocity at age 41. The absolute velocity then decreases gradually to zero from ages 42 to 49. Given that the horizontal location of the centre is to the right of the quadrants of the Cartesian plane, the age distribution of fertility rates is right-skewed. This shape of the phase-plane plot indicates a net positive velocity, where the speed increasing to the maximum fertility rate is faster than the speed decreasing to the minimum fertility rate.

In contrast, the age-specific fertility rates in 2006 show an increase in velocity from ages 15 to 18, and then a decrease in velocity from ages 19 to 21 before another increase in velocity from ages 22 to 26. This indicates the presence of bimodality in the data. The maximum positive velocity is reached at age 26; then the velocity decreases from ages 27 to 30. The maximum fertility occurs at approximately age 31, as shown by zero velocity. From ages 32 to 35, there is an increase in absolute velocity, before reaching the maximum negative velocity at age 36. From ages 37 to 49, the absolute velocity gradually reaches zero. Given that the horizontal location of the centre is to the left of the quadrants of the Cartesian plane, the age distribution of fertility rates is left-skewed. This shape of the phase-plane plot indicates a net negative velocity, where the speed increasing to the maximum fertility rate is slower than the speed decreasing to the minimum fertility rate. Based on the size of the radius of the two circles, the rate of change is greater in 1921 than in 2006. 

Using the animate package \citep{Grahn16} in \LaTeX, we created a dynamic phase-plane plot to visualise systematic changes in age-specific fertility rates over the years. Although such a dynamic plot can be visualised only in the Adobe$^{\copyright}$ pdf file, we present the dynamic phase-plane plot of the Australian age-specific fertility rates observed over the years in the online supplementary material, along with the \textsf{R} code.

\subsubsection{Chilean fertility rates}

\begin{figure}[!ht]
\centering
\subfigure[Phase-plane plot]
{\includegraphics[width=10.5cm]{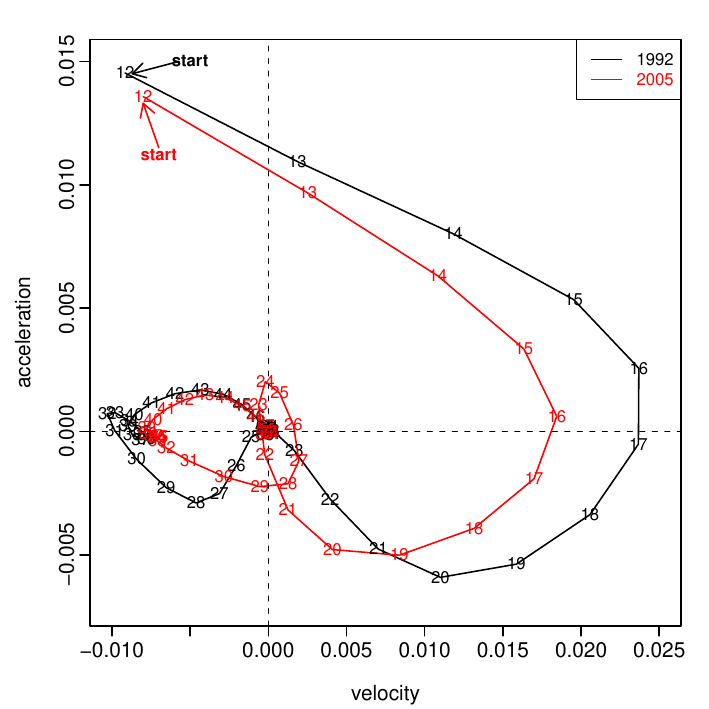}\label{fig:3a_CHL}}
\qquad
\subfigure[Scatter plot]
{\includegraphics[width=10.5cm]{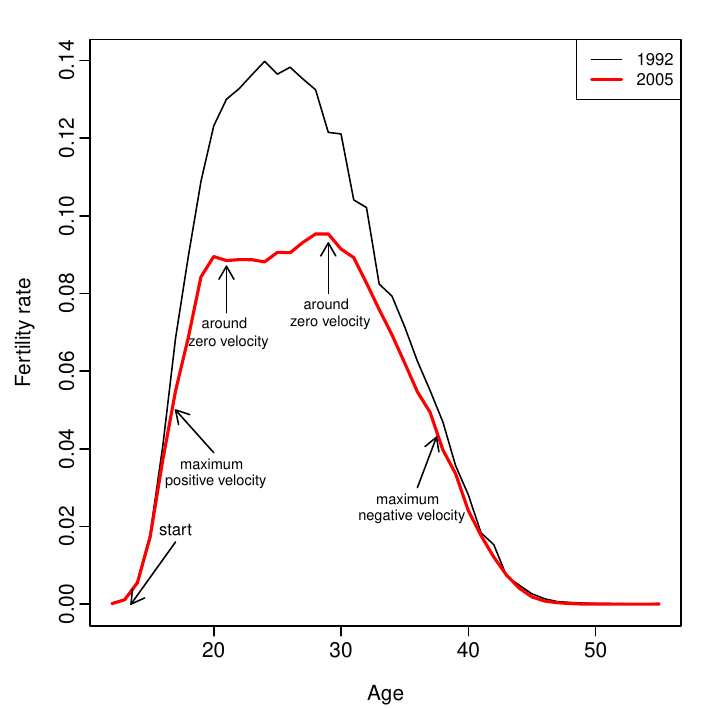}\label{fig:3b_CHL}}
\caption{Phase-plane plot and scatter plot of the age-specific fertility rates for Chile in 1992 (as shown by black colour) and 2005 (as shown by red colour).}\label{fig:3_CHL}
\end{figure}

The phase-plane plot allows the visualisation of changes in the shapes of the cycles from year to year. From these changes in shapes, we aim to reveal social changes in age-specific fertility, including the phenomenon of bimodality \citep{AM11}. Fig.~\ref{fig:3_CHL} present the phase-plane plot and scatter plot for age-specific fertility rates at the first year of data in 1992 and last year of data in 2005 to highlight the specific contrasts between these years in the age distribution of fertility rates.

The age-specific fertility rates in 1992 show an increase in velocity from ages 13 to 16, reaching the highest positive velocity at age 17, before decreasing in velocity from ages 18 to 24. The maximum fertility occurs at age 24, as shown by zero velocity. The absolute velocity increases slowly from ages 25 to 31 and reaches the maximum negative velocity at age 31. The absolute velocity then decreases gradually to zero from ages 32 to 55. Given that the horizontal location of the centre is to the right of the quadrants of the Cartesian plane, the age distribution of fertility rates is right-skewed. This shape of the phase-plane plot indicates a net positive velocity, where the speed increasing to the maximum fertility rate is faster than the speed decreasing to the minimum fertility rate.

In contrast, the age-specific fertility rates in 2005 show an increase in velocity from ages 13 to 16, and then a decrease from age 17 to approximately age 24 before another increase in velocity from ages 25 to 29. This shape of the phase-plane plot indicates the presence of bimodality in the data. The maximum fertility rate occurs at ages 24 and 29, as shown by zero velocity. From ages 30 to 39, there is an increase in absolute velocity, before reaching the maximum negative velocity at approximately age 39. From ages 40 to 55, the absolute velocity gradually reaches zero. Given that the horizontal location of the centre is to the right of the quadrants of the Cartesian plane, the age distribution of fertility rates is right-skewed. This shape of the phase-plane plot indicates a net positive velocity, where the speed increasing to the maximum fertility rate is faster than the speed decreasing to the minimum fertility rate. Based on the size of the radius of the two circles, the rate of change is greater in 1992 than in 2005. 

Using the animate package in \LaTeX, we created a dynamic phase-plane plot to visualise systematic changes in fertility rates over the years. Although such a dynamic plot can be visualised only in Adobe$^{\copyright}$ pdf file, we present the dynamic phase-plane plot of the Chilean age-specific fertility rates observed over the years in the online supplementary material, along with the \textsf{R} code.

\subsection{Comparison of year-specific fertility rates}

\subsubsection{Australian fertility rates}

\begin{figure}[!htbp]
\centering
\subfigure[Phase-plane plot]
{\includegraphics[width=10.3cm]{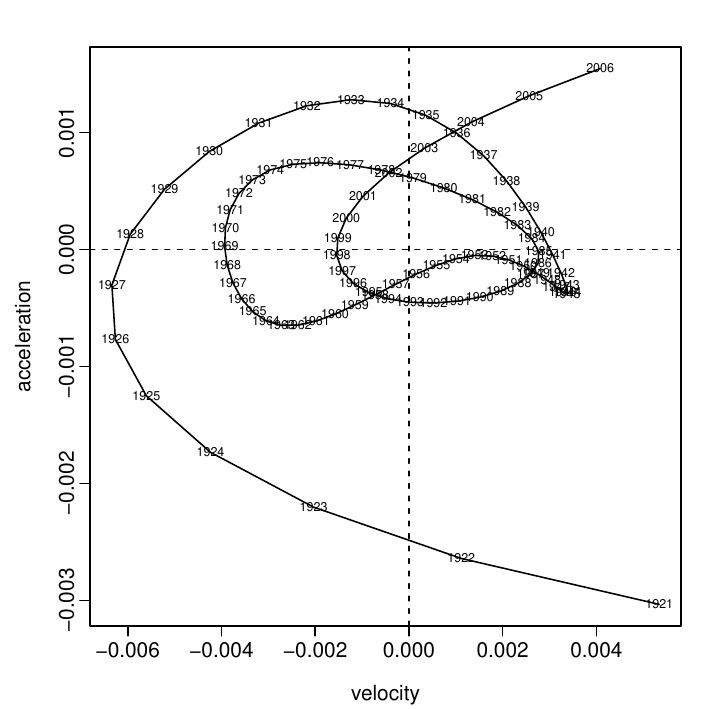}\label{fig:4a}}
\subfigure[Scatter plot]
{\includegraphics[width=10.3cm]{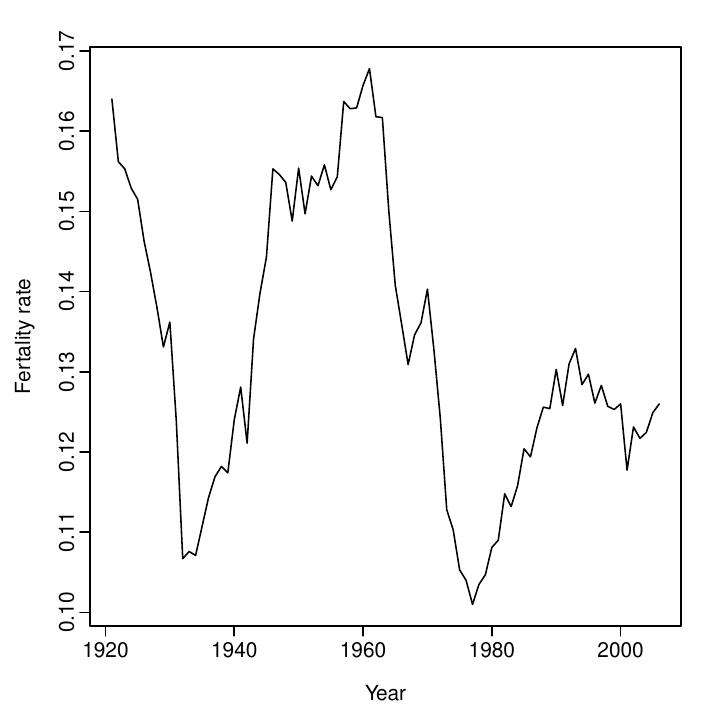}\label{fig:4b}}
\caption{Phase-plane plot and scatter plot of the Australian fertility rates at age 30 observed from 1921 to 2006.}\label{fig:4}
\end{figure}

In Fig.~\ref{fig:4}, we present the phase-plane plot and scatter plot for the Australian fertility rates at a given age, for example, age 30 observed over the years. The phase-plane plot demonstrates a decreasing velocity (i.e., a decreasing trend of fertility rates) from 1921 to 1927, the age-specific fertility rate reaches the highest negative velocity at approximately 1927, before showing an increasing acceleration from 1928 to 1933. A local minimum fertility rate occurs at approximately 1934, as shown by zero velocity in the first and second quadrants. The velocity increases slowly from 1935 to 1940 and reaches a maximum positive velocity at approximately 1941. The velocity increases slightly from 1942 to 1944, before decreasing from 1945 to 1956. A maximum fertility rate occurs at 1957, as shown by zero velocity in the third and fourth quadrants. The velocity then decreases from 1957 to 1968, reaching the second highest negative velocity at 1969. The phase-plane plot indicates the presence of bimodality in the data. 

From 1970 to 1978, the velocity gradually decreases (i.e., a decreasing trend of fertility rates), before another minimum fertility rate occurred at 1979. From 1980 to 1984, the velocity again increases, then declines from 1985 to 1992. Another local maximum fertility rate occurred at approximately 1993. The velocity then decreases from 1994 to 1998, reaching the third highest negative velocity at approximately 1999. From 1999 to 2006, the velocity gradually increases with a local minimum fertility rate occurred at around 2003.

Using the animate package in \LaTeX, we created a dynamic phase-plane plot to visualise systematic changes in year-specific fertility rates over the ages. Although such a dynamic plot can be visualised only in the Adobe$^{\copyright}$ pdf file, we present the dynamic phase-plane plot of the Australian year-specific fertility rates over the ages in the online supplementary material, along with the \textsf{R} code.

\subsubsection{Chilean fertility rates}

\begin{figure}[!htbp]
\centering
\subfigure[Phase-plane plot]
{\includegraphics[width=10.3cm]{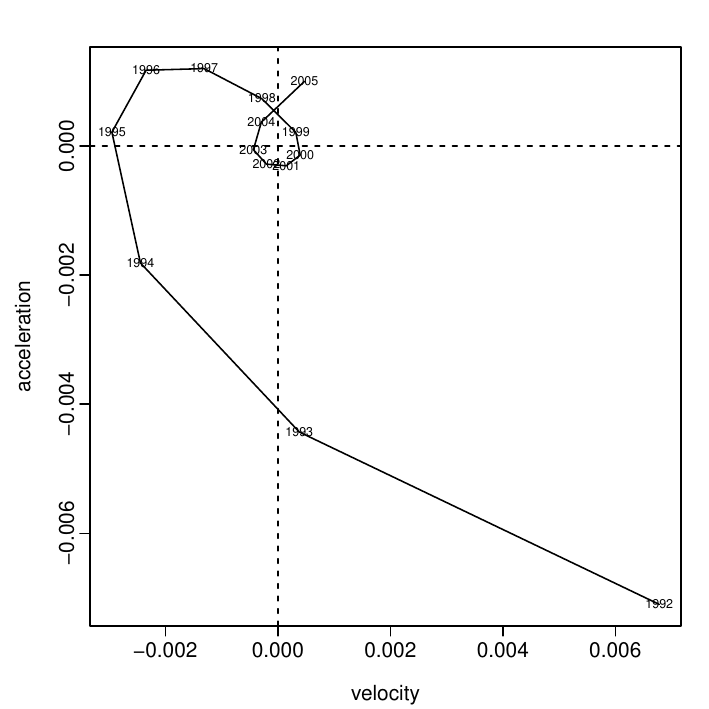}\label{fig:4a_CHL}}
\subfigure[Scatter plot]
{\includegraphics[width=10.3cm]{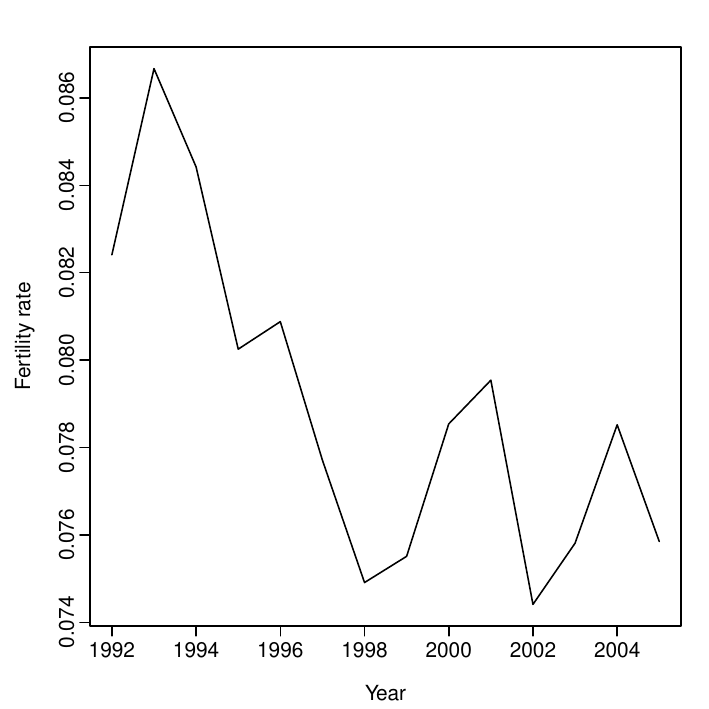}\label{fig:4b_CHL}}
\caption{Phase-plane plot and scatter plot of the Chilean fertility rates at age 33 observed from 1992 to 2005.}\label{fig:5}
\end{figure}

In Fig.~\ref{fig:5}, we present the phase-plane plot and scatter plot for the Chilean fertility rates at a given age, for example, age 33 observed over the years. The phase-plane plot shows a decreasing velocity (i.e., a decreasing trend of fertility rates) from 1992 to 1993, and then an increasing velocity from 1994 to 1995. The maximum negative velocity occurs at 1995, before showing a gradual decrease in absolute velocity from 1996 to 1998. The maximum positive velocity occurs between 1999 and 2000, while the maximum fertility rate occurs between 2001 and 2002, as shown by zero velocity. The absolute velocity decreases from 2003 to 2004, before showing an increase from 2004 to 2005. 

Using the animate package in \LaTeX, we created a dynamic phase-plane plot to visualise systematic changes in year-specific fertility rates over the ages. Although such a dynamic plot can be visualised only in the Adobe$^{\copyright}$ pdf file, we present the dynamic phase-plane plot of the Chilean year-specific fertility rates over the ages in the online supplementary material, along with the \textsf{R} code.

\section{Australian birth-cohort fertility rates}\label{sec:4}

\begin{figure}[!htbp]
\centering
\subfigure[Phase-plane plot]
{\includegraphics[width=10.5cm]{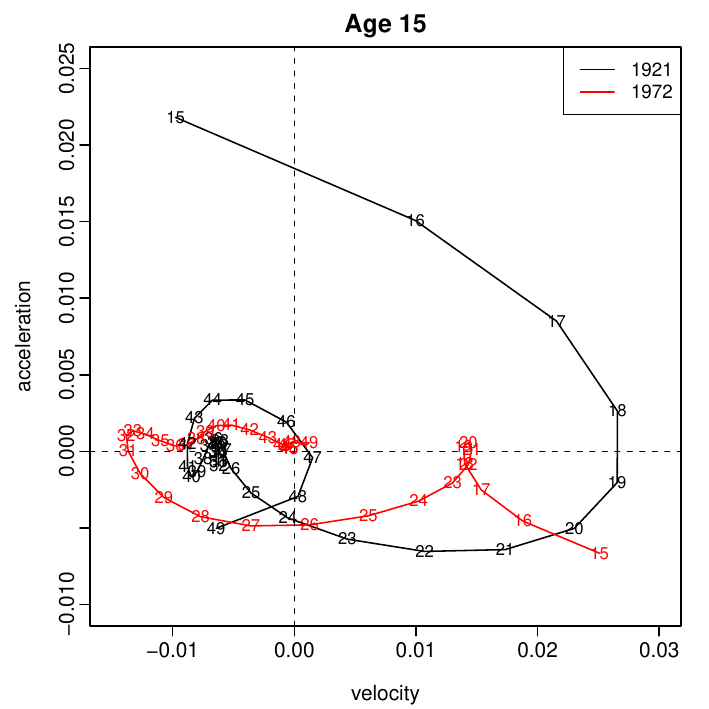}}
\qquad
\subfigure[Scatter plot]
{\includegraphics[width=10.5cm]{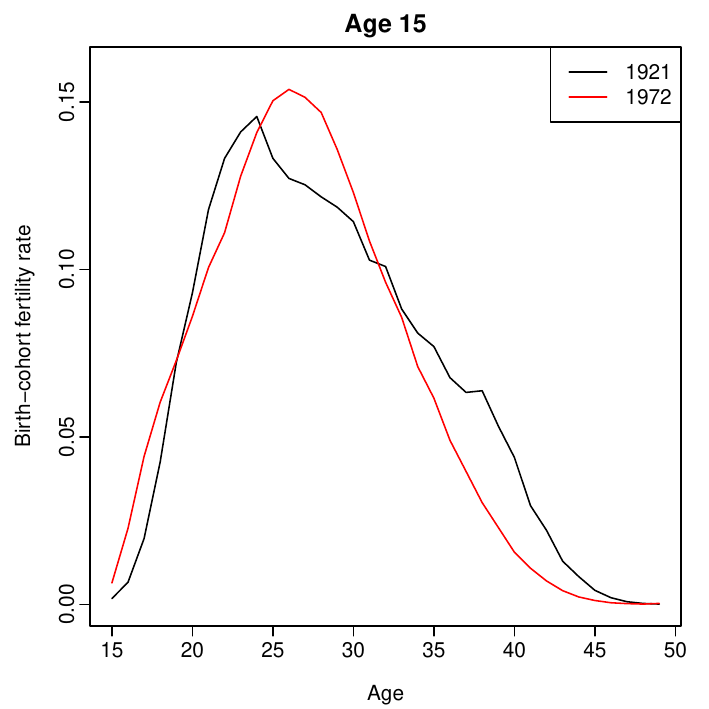}}
\caption{Phase-plane plot and scatter plot of the birth-cohort fertility rates for Australia in 1921 (as shown by black colour) and 1972 (as shown by red colour).}\label{fig:2_cohort}
\end{figure}

In Figure~\ref{fig:2_cohort}, we also consider the phase-plane plot and scatter plot of the Australian birth-cohort fertility rates at an initial age of 15 for years 1921 and 1972, respectively. It reveals that the birth-cohort fertility rates in 1921 increases in velocity from ages 15 to 19, reaching the highest positive velocity at age 19, before showing a decreasing velocity from ages 20 to 23. Maximum fertility occurs at age 24, as shown by zero velocity. The absolute velocity increases slowly from ages 25 to 41 and reaches the maximum negative velocity at age 42. The absolute velocity then decreases gradually to zero from ages 43 to 47, before a slight increase from ages 48 to 49. Given that the horizontal location of the centre is to the right of the quadrants of the Cartesian plane, the age distribution of fertility rates is right-skewed. This shape of the phase-plane plot indicates a net positive velocity, where the speed increasing to the maximum fertility rate is faster than the speed decreasing to the minimum fertility rate.

The birth-cohort fertility rates in 1972 show a decrease in velocity from ages 15 to 18, and then an increase in velocity from ages 19 to 21 before another decrease in velocity from ages 22 to 25. This indicates the presence of bimodality in the data. The maximum fertility occurs at approximately age 26, as shown by zero velocity. From ages 27 to 30, there is an increase in absolute velocity, before reaching the maximum negative velocity at age 31. From ages 32 to 49, the absolute velocity gradually reaches zero.

\section{Conclusion}\label{sec:5}

From the perspective of functional data analysis, we consider the problem of visualisation in demographic research. We first introduce the phase-plane plot to visualise and compare the rate of change in age-specific fertility rates observed over the years. By treating time as a continuous variable, the phase-plane plot can be used to visualise the year-specific fertility rates observed at a given age. Via the animate package, the dynamic phase-plane plot can be applied to visualise rate of change in the Australian and Chilean age-specific fertility rates over years and year-specific fertility rates over the ages. Similarly to the scatter plot and Lexis surface, the phase-plane plot identifies the age associated with the maximum fertility rate and displays the skewness of age-specific fertility distribution based on net velocity. Unlike the scatter plot and Lexis surface, the phase-plane plot identifies the age associated with maximum positive or maximum negative velocity and may be used to compare the magnitude of the rate of change between two years or two ages based on the size of the radius of circles. However, it is possible that two radii of circles may be cumbersome to compare between two consecutive years or ages. We hope that the phase-plane plot will be considered as a complementary graphical tool to the scatter plot and Lexis surface for exploring the rate of change in demographic functions, including age-specific mortality, migration and population size.

As a future work, derivatives can also be useful in predictive data analysis. It is possible to take derivatives of original functions as function-valued explanatory variables in a regression setting \citep[see, e.g.,][]{MP09}.

\section*{Supplemental material}

Supplement to the manuscript titled ``Visualising rate of change: an application to age-specific fertility rates" by H. L. Shang: This supplement contains the four dynamic plots and computational code in \textsf{R} used for producing all the figures. 

\section*{Acknowledgments}

The author acknowledges insightful comments and suggestions from two reviewers and the Associate Editor. Thanks also to Professors Jakub Bijak and Peter W. F. Smith for insightful discussion. 

\newpage

\bibliographystyle{rss}
\bibliography{phase-plane_plot}

\end{document}